\documentclass{llncs} % instead of \documentclass[11pt]{article}
\newtheorem{defn}{D\'efinition} %[section]

\newtheorem{rem}[defn]{Remark}

\title{A Partially Synchronizing  Coloring\thanks{http://www.cs.biu.ac.il/$\sim$trakht}}
\date{}
\author{A.N. Trahtman\thanks{Email: trakht@macs.biu.ac.il}}
\institute{Bar-Ilan University, Dep. of Math., 52900, Ramat Gan,
Israel}

\begin{document}

\maketitle
\centerline{Lecture Notes in Computer Science, 6072(2010), 363-370} 

\begin{abstract}
Given a finite directed graph, a coloring of its edges turns the graph into
a finite-state automaton.
A $k$-synchronizing word of a deterministic automaton is a word in
the alphabet of colors at its edges that maps the state set of the automaton
at least on $k$-element subset. A coloring of edges of a
directed strongly connected finite graph of a uniform outdegree
(constant outdegree of any vertex)  is $k$-synchronizing if the coloring
turns the graph into a deterministic finite automaton possessing a
$k$-synchronizing word.

 For $k=1$ one has the well known road coloring problem.
The recent positive solution of the road coloring problem implies an elegant
generalization considered first by B{\'e}al and Perrin:
a directed finite strongly connected graph of uniform outdegree is
$k$-synchronizing iff the greatest common divisor of lengths of all
 its cycles is $k$.

Some consequences for coloring of an arbitrary finite digraph are
presented.
 We describe a subquadratic algorithm of the road coloring for the
$k$-synchronization implemented in the package TESTAS. A new
linear visualization program demonstrates the obtained coloring.
Some consequences for coloring of an arbitrary finite digraph and of
such a graph of uniform outdegree are presented.

\end{abstract}
{\bf Keywords}: graph, algorithm, synchronization, road coloring, deterministic finite
automaton.
 \section*{Introduction}
The famous road coloring problem was stated almost 40 years ago
\cite{AW} for a strongly connected directed graph of uniform outdegree
where the greatest common divisor (gcd) of lengths of all its cycles is one.

Together with the \v{C}erny conjecture \cite {Ce}, \cite{MS}, the
road coloring problem was once one of the most fascinating
problems in the theory of finite automata. In the popular Internet
Encyclopedia "Wikipedia" it is on the list of most interesting unsolved
problems in mathematics. The recent positive solution of the
road coloring conjecture \cite {Tc}, \cite {Ti}, \cite {Tr} has
posed a lot of generalizations and new problems.

One of them is $k$-synchronizing coloring. A solution of the problem
based on the method from \cite {Tr} was appeared first in \cite{BP}
and repeated later independently in \cite{BF}, \cite{Tk}.

Some consequences for coloring of an arbitrary finite digraph
as well as for coloring of such a graph of a uniform outdegree are a matter
of our interest. The minimal value of $k$ for $k$-synchronizing coloring
exists for any finite digraph and therefore a partially synchronizing
coloring can be obtained.

 Here we describe a polynomial time algorithm for $k$-synchronizing coloring.
Our proof also is based on \cite {Tr} and \cite {Tk}, the more so
the proofs in \cite{BP} meanwhile have some gaps. The theorems and
lemmas from \cite {Tr} are presented below without proof. The
proofs are given only for new or modified results.  The
realization of the algorithm is demonstrated by a new linear
visualization program \cite{BCT}. For an $n$-state digraph with
uniform outdegree $d$, the time complexity of the algorithm is
$O(n^3d)$ in the worst case and $O(n^2d)$ in the majority of
cases. The space complexity is quadratic.

 \section*{Preliminaries}
As usual, we regard a directed graph with colors assigned to its edges as a finite
automaton, whose input alphabet $\Sigma$ consists of these colors.

A directed graph with constant outdegree (the number of outgoing edges)
of all its vertices is called a graph of {\it uniform outdegree}.

A finite directed strongly connected graph of uniform outdegree where
the gcd of lengths of all its cycles is $k$ will be called {\it k-periodic}.

An automaton is {\it deterministic} if no state has two outgoing edges
of the same color. In a {\it complete} automaton each state has outgoing
edges of any color.

If there exists a path in an automaton from the state $\bf p$ to
the state $\bf q$ and the edges of the path are consecutively
labelled by $\sigma_1, ..., \sigma_k \in \Sigma$, then for
$s=\sigma_1...\sigma_k \in \Sigma^+$ let us write  ${\bf q}={\bf
p}s$.

Let $Ps$ be the map of the subset $P$ of states of an automaton using
the word $s \in \Sigma^+$. For the transition graph
$\Gamma$ of an automaton let $\Gamma s$
denote the map of the set of states of the automaton.

 Let $|P|$ denote the size of the subset $P$ of states from an automaton
  (of vertices from a graph).

 A word $s \in \Sigma^+ $ is called a {\it $k$-synchronizing}
word of the automaton with transition graph $\Gamma$
if both $|\Gamma s|=k$ and for all words $t\in \Sigma^*$ holds $|\Gamma t|\ge k$.

 A coloring of a directed finite graph is {\it $k$-synchronizing} if the
coloring turns the graph into a deterministic finite automaton
possessing a $k$-synchronizing word and the value of $k$ is minimal.

  A pair of distinct states $\bf p, q$ of an automaton (of vertices of
  the transition graph)  will be called  {\it synchronizing} if
  ${\bf p}s = {\bf q}s$ for some $s \in \Sigma^+$.
In the opposite case, if for any $s$ ${\bf p}s \neq {\bf q}s$, we
call the pair {\it deadlock}.

A synchronizing pair of states $\bf p$, $\bf q$ of an automaton is
called {\it stable} if for any word $u$ the pair of states ${\bf p}u, {\bf
q}u$ is also synchronizing \cite{CKK}, \cite{Ka}.

We call the set of all outgoing edges of a vertex a {\it bunch} if
 all these edges are incoming edges of only one vertex.

The subset of states (of vertices of the transition
graph $\Gamma$) of maximal size such that every pair of states
from the set is a deadlock will be called an {\it $F$-clique}.

\section{A $k$-synchronizing coloring}
 \begin{lemma}  \label {ne}
Let a finite directed strongly connected graph $\Gamma$ with uniform
outdegree have a $k$-synchronizing coloring. Then the greatest common
divisor $d$ of lengths of all its cycles is not greater than $k$.

If the length of a path from the state $\bf p$ to the state $\bf q$
is equal to $i \ne 0$ (modulo $d$) then for any word $s$ one has
${\bf p}s \ne {\bf q}s$.
 \end{lemma}
 Proof.
Let $N$ be the function defined on the states of $\Gamma$ in the following way -
we take an arbitrary vertex $\bf p$ and let $N({\bf p})=0$.
Then for each vertex $\bf q$ with defined $N({\bf q})$ suppose for any
next nearest vertex $\bf r$ $N({\bf r}) = N({\bf q})+1$ (modulo $d$).
The graph is strongly connected, whence for each vertex the function $N$
is defined. The enumeration does not imply a contradiction
anywhere because the length of each cycle is divided by $d$.
Then by any coloring the mapping by a word $t$ produced the same
shift of size $|t|$ (modulo $d$) of the function $N$ on the states.
Therefore the states with distinct values of $N$ could not have common
image and $|\Gamma t| \geq d$ for any word $t$. $|\Gamma s|=k$
for $k$-synchronizing word $s$. Consequently, $k \geq d$.

By any coloring, the mapping by a word $s$ produced the same
shift of the function $N$ on the set of states.
$N({\bf p}s)=N({\bf p})+|s|$ (modulo $d$).
Therefore the difference of the values of the function $N$ on two states
is not changed by the shift.

 \begin{theorem}  \label {ck} \cite{CKK}, \cite{Ka}, \cite{Tc}
Let us consider an arbitrary coloring of a strongly connected
graph $\Gamma$ with constant outdegree.
Stability of states is a binary relation on the set of states
of the obtained automaton. Denote the reflexive closure of this
relation by $\rho$.
Then $\rho$ is a congruence relation, $\Gamma/\rho$ presents a
directed strongly connected graph with constant outdegree, the gcd
$d$ of all its cycles is the same as in $\Gamma$, and a
$k$-synchronizing coloring of $\Gamma/\rho$ implies
a $k$-synchronizing recoloring of $\Gamma$.
 \end{theorem}

     \begin{lemma}  \label {f3} \cite{Tc}
Let $F$ be an $F$-clique via some coloring of a strongly connected graph $\Gamma$.
For any word $s$ the set $Fs$ is also an $F$-clique and each state
[vertex] $\bf p$ belongs to some $F$-clique.
 \end{lemma}
 \begin{lemma}  \label {f6}
Let $A$ and $B$ ($|A|>1$) be distinct $F$-cliques via
some coloring  without stable pairs of the $k$-periodic graph $\Gamma$.
 Then $|A| -|A \cap B| =|B| -|A \cap B| >1$.
 \end{lemma}
Proof. Let us assume the contrary: $|A| -|A \cap B|=1$.
By the definition of $F$-clique, $|A|=|B|$ and so $|B| -|A \cap B|=1$, too.
Thus $|A| -|A \cap B| =|B| -|A \cap B|=1$.

 The pair of states ${\bf p} \in A \setminus B$ and
${\bf q} \in B \setminus A$ is not stable. Therefore for some word
$s$ the pair $({\bf p}s, {\bf q}s)$ is a deadlock.
Any pair of states from the $F$-clique $A$ and from the $F$-clique $B$
as well as from $F$-cliques $As$ and $Bs$ is a deadlock. So each  pair
of states from the set $(A \cup B)s$ is a deadlock, whence $(A \cup B)s$
is an $F$-clique.

One has $|(A \cup B)s| = |A| + 1 > |A|$ in spite of maximality of the size of
$F$-clique $A$.
\begin{lemma}  \label {f7}  \cite{Tc}
Let some vertex of a directed complete graph $\Gamma$ have two incoming bunches.
 Then any coloring of $\Gamma$ has a stable pair.
 \end{lemma}
Proof. If a vertex ${\bf p}$ has two incoming bunches from vertices ${\bf
q}$ and ${\bf r}$, then the pair ${\bf q}$, ${\bf r}$ is stable
for any coloring because ${\bf q}\alpha ={\bf r}\alpha =\bf p$ for
any letter (color) $\alpha \in \Sigma$.
\subsection{The spanning subgraph with maximum of edges in the cycles}
\begin{definition}  \label {d1}
Let us call a subgraph $S$ of the $k$-periodic graph $\Gamma$ a {\emph spanning
subgraph} of $\Gamma$ if all vertices of $\Gamma$ belong to $S$ and exactly one
outgoing edge of every vertex
(in the usual graph-theoretic terms it is 1-outregular spanning subgraph).

 A maximal subtree of the spanning subgraph $S$ with root on a cycle from $S$ and
having no common edges with cycles from $S$ is called a {\emph tree} of $S$.

The length of the path from a vertex ${\bf p}$ through the edges of its tree of the spanning
set $S$ to the root of the tree is called the {\emph  level} of ${\bf p}$ in $S$.

A tree with vertex of maximal level is called a {\emph maximal tree}.
\end{definition}
\begin{rem}  Any spanning subgraph $S$ consists of disjoint cycles
and trees with roots on cycles. Each tree and cycle of $S$ is defined
identically. The level of the vertices belonging to some cycle is zero.
The vertices of the trees except the roost have a positive level. The
vertices of maximal positive level have no incoming edge in $S$.
The edges labelled by a given color defined by any coloring form
a spanning subgraph. Conversely, for each spanning subgraph, there exists
a coloring and a color such that the set of edges labelled with this
color corresponds to this spanning subgraph.
\end{rem}
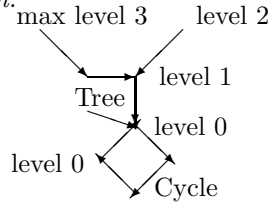
\begin{picture}(170,68)
\put(127,54){\vector(0,-1){18}}
\put(145,72){\vector(-1,-1){18}}
\put(109,54){\vector(1,0){18}}
\put(91,72){\vector(1,-1){18}}
\put(127,36){\vector(1,-1){15}}
\put(114,23){\vector(1,1){15}}
\put(127,10){\vector(-1,1){15}}
\put(140,23){\vector(-1,-1){15}}
\put(82,74){max level 3}
\put(80,18){level 0}
\put(134,9){Cycle}
\put(150,74){level 2}
\put(136,51){level 1}
\put(134,32){level 0}
\put(104,43){Tree}
\put(109,41){\vector(3,-1){19}}
 \end{picture}
\begin{lemma}  \label {p1}
Let $N$ be a set of vertices of level $n$ from some tree of the
spanning subgraph $S$ of the $k$-periodic graph $\Gamma$. Then via a
coloring of $\Gamma$ such that all edges of $S$ have the same
color $\alpha$, for each $F$-clique $F$ holds $|F \cap N| \leq 1$.
 \end{lemma}
Proof. Some power of $\alpha$ synchronizes all states of a given
level of the tree and maps them into the root. Each pair of
states from an $F$-clique could not be synchronized and therefore
could not belong to $N$.

\begin{lemma}  \label {f8}  \cite{Tc}
Let $d$-periodic graph $\Gamma$ have a spanning subgraph $R$
consisting solely of disjoint cycles (without trees).
 Then $\Gamma$ either is a cycle of length $d$ of bunches or has another spanning
subgraph with exactly one maximal tree.
\end{lemma}
 \begin{lemma}  \label {f9} \cite{Tr}
Assume that no vertex of the graph $\Gamma$ has two incoming bunches.
Let $R$ be a spanning subgraph with non-trivial tree and let its
 tree $T$ with the root $\bf r$ on cycle $H$ have all vertices
of maximal level $L$ and one of them is the vertex $\bf p$.
Let us consider

1) replacing an edge $e$ from $R$ with an edge having the same start vertex as
$e$ and ending in $\bf p$,

 2) replacing in $R$ an incoming edge of a root on the path in the tree from $\bf p$,

3) replacing in $R$ an incoming edge of a root from $H$.

Suppose that at most two such operations do not increase the number
of edges in cycles. Then by the coloring of $R$ by the color $\alpha$ $\Gamma$
 has a spanning subgraph with one maximal tree and the pair
${\bf p}\alpha^{L-1}$,  ${\bf p}\alpha^{L+|H|-1}$ is stable.
 \end{lemma}
\begin{theorem}  \label {t2}
 Any $k$-periodic graph $\Gamma$ of size greater than $k$ has a
coloring with stable pair.
 \end{theorem}
Proof. We have $|\Gamma|>k$. By Lemma \ref{f7}, in the case of vertex
with two incoming bunches  $\Gamma$ has a coloring with stable pairs.
 In opposite case, by
 Lemmas \ref{f9} and \ref{f8},  $\Gamma$ has a spanning subgraph $R$ with
 one maximal tree in $R$.

Let us give to the edges of $R$ the color $\alpha$
and denote by $C$ the set of all vertices from
the cycles of $R$.
Then let us color the remaining edges of $\Gamma$ by other colors
arbitrarily.

By Lemma \ref{f3}, in a strongly connected graph $\Gamma$ for
every word $s$ and $F$-clique $F$ of size $|F| > 1$, the set $Fs$
also is an $F$-clique of the same size and for
any state $\bf p$ there exists an $F$-clique $F$ such that ${\bf
p} \in F$.

In particular, some $F$ has non-empty intersection with the set
$N$ of vertices of maximal level $L$. The set $N$ belongs to one
tree, whence by Lemma \ref{p1} this intersection has only one
vertex.
 The word $\alpha^{L-1}$ maps $F$ on an $F$-clique $F_1$ of size
$|F|$. One has $|F_1 \setminus C|=1$ because the sequence of edges of
the color $\alpha$ from any tree of $R$ leads to the root of
the tree, and the root belongs to a cycle colored by $\alpha$ from $C$ and
 only for the set $N$ with vertices of maximal level holds
 $N\alpha^{L-1} \not\subseteq C$. So
 $|N\alpha^{L-1}  \cap F_1|=|F_1 \setminus C|=1$ and
$|C \cap F_1|=|F_1|-1$.

Let the integer $m$ be a common multiple of the lengths
of all considered cycles from $C$ colored by $\alpha$.
 So for any $\bf p$ from $C$ as well as from $F_1 \cap C$
holds ${\bf p}\alpha^m={\bf p}$. Therefore for an $F$-clique
$F_2=F_1\alpha^m$ holds $F_2 \subseteq C$ and
$C \cap F_1 =F_1 \cap F_2$.

Thus two $F$-cliques $F_1$ and $F_2$ of size $|F_1|>1$ have
$|F_1|-1$ common vertices. So $|F_1 \setminus (F_1 \cap F_2)|=1$.
Consequently, in view of Lemma \ref{f6}, there exists a stable
pair in the considered coloring.
\begin{theorem}  \label {t}
Every strongly connected graph $\Gamma$ is $k$-periodic if and only if
the graph has a $k$-synchronizing coloring.
For arbitrary coloring $\Gamma$ is $m$-synchronizing for $m\geq k$.
 \end{theorem}
Proof.
 From the $k$-synchronizing coloring of $\Gamma$ by Lemma \ref{ne} follows that
if $\Gamma$ is $m$-periodic graph then $m \geq k$. So $|\Gamma| \geq k$.
If $|\Gamma|=k$ then also $m=k$.

Thus we only need to consider the case $|\Gamma|>k$. By  Lemmas \ref{f8} and \ref{f9}
there exists a spanning subgraph with one maximal tree, whence by Theorem \ref{t2}
there exists a stable pair of states. Theorem \ref{ck} reduced the problem
to a homomorphic image of $\Gamma$ of smaller size and the induction finishes the proof.
\\
\\
The theorem \ref{t} implies some consequences, in particular, even for an arbitrary digraph.
\begin{lemma}  \label{cog}
Let a finite directed graph $\Gamma$ have a sink component $\Gamma_1$. Suppose that
by removing some edges of $\Gamma_1$ one obtains strongly connected
 directed graph $\Gamma_2$ of uniform outdegree.  Let $k$ be the greatest common divisor
of lengths of the cycles of $\Gamma_2$.
Then $\Gamma$ has $k$-synchronizing coloring.
 \end{lemma}
Proof. $\Gamma_2$ has a $k$-synchronizing coloring by Theorem \ref{t}. The edges outside
$\Gamma_2$ can be colored arbitrarily, sometimes by adding new colors for to obtain
deterministic automaton.
\begin{lemma}  \label{co1}
Let $\Gamma_1$,..., $\Gamma_m$ be strongly connected components of a
finite directed graph $\Gamma$ of uniform outdegree such that no edge
leaves the component $\Gamma_i$ ($0< i \leq m$). Let $k_i$ be the greatest common divisor
of lengths of cycles in $\Gamma_i$ and suppose $k=\sum_{j=1}^m k_j$.
Then $\Gamma$ has $k$-synchronizing coloring.
 \end{lemma}
The Lemmas \ref{cog}, \ref{co1} imply
\begin{theorem}  \label {t4}
Let finite directed graph $\Gamma$ have a subgraph of uniform outdegree
with strongly connected components $\Gamma_1$,..., $\Gamma_m$ such that no edge
leaves the component $\Gamma_i$ ($0< i \leq m$) in the subgraph.
Let $k_i$ be the greatest common divisor
of lengths of cycles in $\Gamma_i$ and suppose $k=\sum_{j=1}^m k_j$.
Then $\Gamma$ has a $k$-synchronizing coloring.
 \end{theorem}
The proof for arbitrary $k$ did not use anywhere that $k \ne 1$, whence from
the validity of the Road Coloring Conjecture we have the following:
\begin{theorem}  \label{t5} \cite{Tr} \cite{Tc}
Every finite strongly connected graph $\Gamma$ of uniform outdegree with greatest common
divisor of all its cycles equal to one has synchronizing coloring.
 \end{theorem}
\section{Find minimal $k$ for $k$-synchronizing coloring}
The algorithm is based on Theorem \ref{t} and Lemma \ref{cog}.
 One must check the existence of strongly connected components $(SCC)$
having no outgoing edges to others $SCC$ and check the condition on gcd in
any such $SCC$ $H$. There exists a subgraph $S$
of $C$ of maximal uniform outdegree. Then we check the condition on
gcd in any such $S$. Let us use the linear algorithm of finding
strongly connected component $SCC$ \cite{AHU}. Then we mark all $SCC$
having outgoing edges to others $SCC$ and study only all remaining $SCC$ $H$.

The coloring of $H$:
We find in every $H$ the value of $k_H$ for the
$k_H$-synchronizing coloring of $H$.  Let $\bf p$ be a vertex from $H$. Suppose
$d({\bf p})=1$. For an edge ${\bf r} \to {\bf q}$ where $d({\bf r})$ is
already defined and $d({\bf q})$ is not defined suppose
$d({\bf q})=d({\bf r})+1$. If $d({\bf q})$ is defined let us add
the difference $abs(d({\bf q})-1-d({\bf r}))$ to the set $D$ and
count the gcd of the integers from $D$. If finally $gcd=k_H$, then the $SCC$
 $H$ has $k_H$-synchronizing coloring. The value of $k$ is equal to the sum
of all such $k_H$ (Theorem \ref{t}). The search of $k$ is linear.

The edges outside subgraphs $H$ can
 be colored arbitrarily (sometimes by additional colors).
The outgoing edges of every state are colored by different colors.

\section{The algorithm for $k$-synchronizing coloring}
Let us begin from an arbitrary coloring of a directed graph $\Gamma$ with
$n$ vertices and $d$ outgoing edges of any vertex. The considered
$d$ colors define $d$ spanning subgraphs of the graph. We find all
$SCC$ $H$ having no outgoing edges and study every such $SCC$
$H$ separately.

We keep images of vertices and colored edges from the generic graph by
any transformation and homomorphism.

If there exists a loop in $\Gamma$ then let us color the edges of
a tree with root in the vertex of a loop by one color. The other
edges may be colored arbitrarily. The coloring in this case is
synchronizing for the considered $SCC$.

In the case of two incoming bunches of one vertex the beginnings of
these bunches form a stable pair by any coloring (Lemma \ref{f7}).
We merge both vertices in the homomorphic image of the graph
(Theorem \ref{ck}) and obtain according to the theorem a new graph
of the size less than $|\Gamma|$.

The linear search of two incoming bunches and of loop can be made at any stage
of the algorithm.

Find the parameters of the spanning subgraph: levels of
all vertices, the number of vertices (edges) in cycles, for any
vertex let us keep its tree and the cycle of the root of the tree.
We form the set of vertices of maximal level and choose from the
set of trees
 a tree $T$ with vertex $\bf p$ of maximal level. This step is linear
 and used by any recoloring.
\begin{lemma}  \label {al}
Let graph $\Gamma$ have two cycles $C_u$ and $C_v$ either with one common vertex
${\bf p}_1$ or with a common sequence ${\bf p}_1$,..., ${\bf p}_k$, such that
all incoming edges of ${\bf p}_i$ form a bunch from ${\bf p}_{i+1}$ ($i<k$).
 Let $u \in C_u$ and $v \in C_v$ be the edges of the cycles leaving ${\bf p}_1$.
Let $T$ be a maximal subtree of $\Gamma$ with the root ${\bf p}_1$ and edges from $C_u$ and
$C_v$ except $u$ and $v$.

Then the the adding one of the edges $u$ or $v$ turns the subtree $T$ into precisely one maximal
 tree of the spanning subgraph.
  \end{lemma}
Proof.
 Let us add to $T$ either $u$ or $v$
and then find the maximal levels of vertices in both cases.
The vertex ${\bf p}_i$ for $i>1$ could not be the root of a tree.
If any tree of spanning subgraph with vertex of maximal level has
the root ${\bf p}_1$ then in both cases the lemma holds.
If some tree of spanning subgraph with vertex of maximal level has the
root only on $C_u$ then let us choose the adding of $v$.
In this case the level of the considered vertex is growing and
only the new tree with root ${\bf p}_1$ has vertices of maximal level.
In the case of the root on $C_v$ let us add $u$.
\\
\\
1) If there are two cycles with one common vertex then we use
Lemma \ref{al} and find a spanning subgraph $R_1$ such that any
vertex $\bf p$ of maximal level $L$ belongs to one tree with root
on a cycle $H$. Then after coloring edges of $R_1$ with the color $\alpha$
we find stable pair ${\bf q}= {\bf p}\alpha^{L-1+|H|}$ and
${\bf s}=\bf p\alpha^{L-1}$ (Lemma \ref{f9}) and go to step 3).
The search of a stable pair in this case is linear and the whole algorithm
therefore is quadratic.

2) Let us consider now the three replacements from Lemma \ref{f9} and find
the number of edges in cycles and other parameters of the spanning subgraph
of the given color. If the number of edges in the cycles is growing, then the new
spanning subgraph must be considered and the new parameters of the subgraph
must be found. In the opposite case, after at most $3d$ steps,
by Lemma \ref{f9}, there exists a tree $T_1$ with root on the cycle $H_1$ of
a spanning subgraph $R_1$ such that any vertex $\bf p$ of maximal level $L$ belongs
to $T_1$.

Suppose the edges of $R_1$  are colored by the color $\alpha$.
Then the vertices ${\bf q}= {\bf p}\alpha^{L-1+|H_1|}$ and
${\bf s}=\bf p\alpha^{L-1}$ by Lemma \ref{f9} form a stable pair.

3) Let us finish the coloring and find the subsequent stable pairs of the
pair (${\bf s}$, ${\bf q}$) using appropriate coloring. Then we go to the
homomorphic image $\Gamma_i/\rho$ (Theorem \ref{ck}) of the considered graph
$\Gamma_i$ ($O(|\Gamma_i|m_id)$ complexity where $m_i$ is the size of the
 map $\Gamma_i$).
 Then we repeat the procedure with a new graph $\Gamma_{i+1}$ of a smaller size.
So the overall complexity of this step of the algorithm is
$O(n^2d)$ in the majority of cases and $O(n^3d)$ if the number of edges in
cycles grows slowly, $m_i$ decreases also slowly, loops do not appear and
the case of two ingoing bunches rarely emerges (the worst case).

Let $\Gamma_{i+1} =\Gamma_i/\rho_{i+1}$ on some stage $i+1$ have
$k$-synchronizing coloring. For every stable pair ${\bf q}, {\bf p}$
of vertices from $\Gamma_{i}$ there exists a pair of corresponding
outgoing edges that reach either another stable pair or one vertex.
This pair of edges is mapped on one image edge of $\Gamma_{i+1}$.
 So let us give the color of the image to preimages and obtain in
this way a $k$-synchronizing coloring of $\Gamma_{i}$.  This step
is linear. So the overall complexity of the algorithm is $O(n^3d)$
in the worst case. If Lemma \ref{al} is applicable, then the complexity
is reduced at this branch of the algorithm (as well as at some other branches) to
$O(n^2d)$. The practical study of the implementation of the algorithm demonstrates
mainly $O(n^2d)$ time complexity.

\section{Conclusion}
In this paper, we have continued the study of $k$-synchronizing  coloring
of a finite directed graph. It is an important and natural
generalization of the well known notion of the synchronizing
coloring. This new approach gives an opportunity to extend
essentially the class of studied graphs, removing the restriction
on the size of  the great common divisor of the lengths of the
cycles of the graph. The restriction is inspired  by the Road
Coloring Problem and now for $k$-synchronizing  coloring can be
omitted.

There exists still another restriction on the properties of the
considered graphs concerning the uniform outdegree of the vertex.
However, it also can be omitted by consideration of subgraphs of
the graph (Corollary \ref{cog}).

The practical use of this investigation needs an algorithm and the paper
presents a corresponding polynomial time  algorithm of the road coloring
for the $k$-synchronization implemented in the package TESTAS.
The realization of the algorithm demonstrates a new linear
visualization program \cite{BCT}.

 \end{document}